# Self-referenced Spectral Interferometry for Single-shot Characterization of Ultrashort Free-electron Laser Pulses

Yaozong Xiao[1,2], Hao Sun[1,2], Bo Liu[1,3], Zhentang Zhao[1,3], Chao Feng[1,3] *

[1]*Shanghai Institute of Applied Physics, Chinese Academy of Sciences, Shanghai 201800, China*

[2]*University of Chinese Academy of Sciences, Beijing 100049, China*

[3]*Shanghai Advanced Research Institute, Chinese Academy of Sciences, Shanghai 201210, China*

*Corresponding author. Email: fengc@sari.ac.cn

**Abstract**

Attosecond x-ray pulse with known spectro-temporal information is an essential tool for the investigation of ultrafast electron dynamics in quantum systems. Ultrafast free-electron lasers (FELs) have the unique advantage on unprecedented high-intensity at x-ray wavelengths. However, no suitable method has been established so far for the spectro-temporal characterization of these ultrashort x-ray pulses. In this letter, a simple method has been proposed based on self-referenced spectral interferometry for reconstructing ultrashort FEL pulses both in temporal and spectral domains. We have demonstrated that the proposed method is reliable to reconstruct the temporal profile and phase of attosecond x-ray FEL pulses with an error of only a few-percent-level. Moreover, the first proof-of-principle experiment has been performed to achieve single-shot spectro-temporal characterization of ultrashort pulses from a high-gain FEL. The precision of the proposed method will be enhanced with the decrease of the pulse duration, paving a new way for complete attosecond pulse characterization at x-ray FELs.

*Introduction.* Exploring the elementary processes driving the transformation of matter, such as photoemission delay [1], tunneling delay time [2], valence electron motion in atoms [3], charge migration [4] and proton dynamics [5] in molecules, urgently needs advanced techniques with attosecond temporal resolution and angstrom spatial resolution. The discovery of high-harmonic generation (HHG) [6–10] sets the basis for the generation of attosecond pulses and has opened the opportunity to investigate electronic ultrafast processes on the attosecond time scale. In parallel with the development of the HHG technique, the last decades witness the rise of high-gain x-ray free-electron lasers (XFELs) that have unique advantages in producing ultrashort x-ray pulses with unprecedented peak brightness [11,12]. Various methods [13–21] based on XFELs have been proposed and demonstrated in recent years to further shorten the pulse duration to attosecond regime while maintain the peak power at tens of gigawatts level, complementing state-of-the-art HHG techniques with limited pulse energies at x-ray wavelengths. On the other hand, the characterization of temporal information of these ultrashort XFEL pulses is equally important, because precise knowledge of the spectro-temporal distribution is beneficial for ultrafast experiments designed to make full use of the laser-like properties of XFELs.

The temporal structure of XFEL pulses can be indirectly inferred by measuring the longitudinal phase space of the electron beam after lasing with an x-band radiofrequency transverse deflector [22]. This method has been widely adopted by XFEL facilities and achieved a temporal resolution of a few femtoseconds [23–26]. Recently, direct experimental determinations of the temporal structures of FEL pulses have been accomplished by the terahertz- or laser-assisted streaking measurements with the resolution from femtosecond to attosecond [27–30]. In particular, angular streaking, as a reliable attosecond measurement method, does not require precise knowledge of the arrival time between x-ray pulses and streaking lasers, since the temporal profiles of x-ray pulses can be retrieved from the relative change in angular distribution [31,32]. In general, above methods are only capable for extracting the temporal profiles of XFEL pulses, no method has been established yet to achieve the complete spectro-temporal characterization of attosecond x-ray pulses at present.

Spectral phase interferometry for direct electric-field reconstruction (SPIDER) [33] has been demonstrated to be a reliable technique for spectro-temporal characterization of femtosecond optical lasers using direct data inversion [34,35]. This method is also feasible for the seeded free-electron laser, where a pair of femtosecond pulses with spectral shear could be generated by properly setting the parameters of seed lasers and electron beams [36]. However, for structured spectrum, SPIDER takes on uncertainties for different filter widths [37]. To overcome this problem, the Wavelet-transform (WT) algorithm had been proposed to extract the spectral phase instead of the Fourier-transform algorithm [37,38].

In this letter, based on the WT algorithm, we proposed a novel technique to retrieve the spectro-temporal properties of ultrashort FEL pulses. With an electron beam initiating coherent radiation successively in the main FEL amplifier and an afterburner, a pair of spectrally sheared pulses with an appropriate time delay could be produced, resulting in a self-referenced spectral interferogram (SSI) that can be used for reconstruction of the FEL pulses. The proposed method is suitable for various ultrashort XFEL generation techniques either based on self-amplified spontaneous emission (SASE) or external seeding and will bring a new area of attoscience with high intensity XFELs.

*Methods and 3-D numerical simulations.* The schematic layout of the proposed technique is shown in Fig. 1, where a small chicane and a short undulator (afterburner) has been added downstream of the main FEL amplifier. After generating attosecond x-ray pulse (sample pulse) in the main amplifier, the electron beam with strong microbunching that formed in the main amplifier will initiate coherent radiation (reference pulse) in the following afterburner. The chicane is used to delay the electron beam and inducing a temporal shift between these two radiation pulses. For a sufficiently homogeneous electron beam with constant energy in the lasing fraction, the attosecond pulses generated in the main amplifier and afterburner will have equal temporal phases that follow the distribution of the microbunchings. In the proposed method, the generation of a pair of frequency-sheared pulses is an imperative prerequisite for the phase retrieval procedure. The required slight frequency shift can be induced by the frequency-pulling effect, when a mismatch between the frequency of the microbunching and the resonance of the undulator exists in the afterburner [39]. The wavelength shift can be calculated by:

$$\Delta\lambda = \Delta\lambda_u \sigma_b^2/(\sigma_b^2 + \sigma_u^2). \tag{1}$$

Where $\lambda_u$ is the undulator resonance wavelength, $\sigma_b$ and $\sigma_u$ are the FWHM bandwidths of the microbunching and gain spectra of the afterburner, respectively.

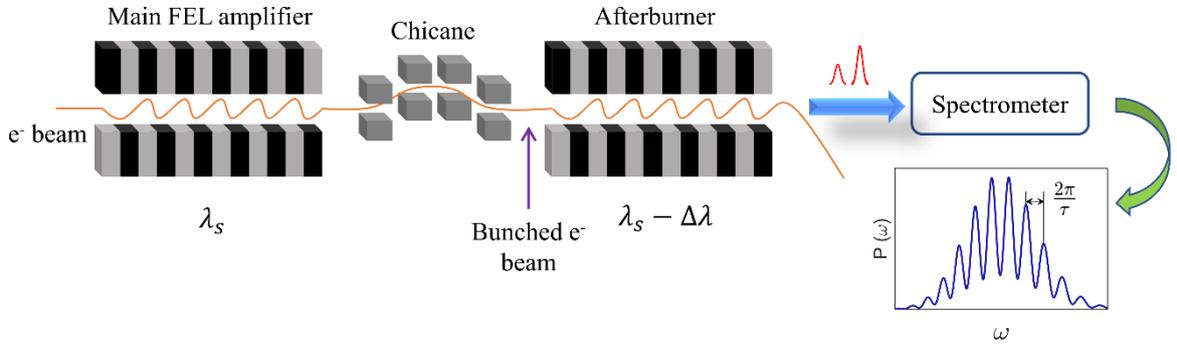

FIG. 1. Schematic layout of the proposed method.

To investigate the performance of the proposed method, three-dimensional simulations with *GENESIS 1.3* [40] have been performed with realistic parameters of the Shanghai soft x-ray FEL facility (SXFEL) [41], as listed in Table I. Based on the method in Ref. [18], attosecond soft x-ray pulses with the central wavelength $\lambda_s$ of 2 nm, peak power of 282 MW and pulse duration of 610 attoseconds were generated in the main undulator with period of 1.6 cm and K value of 1.1565, as shown in Fig. 2(a). After that, the bunched electron beams were sent through a small chicane with a dispersion strength of $R56 \approx 0.9\ \mu m$ to introduce a time delay of $\tau = 1.5\ fs$. Simulations results show that this dispersion has little influence on the microbunchings in the electron beams, where the maximal bunching factor changes from 0.33 to 0.35. Then the electron beam travels through a short afterburner with period of 1.6 cm (same as the main undulator) and length of 0.3 m. The K value of the afterburner was tuned to 1.1559 to slightly shift the central frequency by $\Omega = 3 \times 10^{14}$ rad, fitting perfectly with the calculation results from Eq. 1. A reference pulse with peak power of 92 MW and pulse duration of 560 attoseconds has been generated, as shown in Fig. 2(a). Fig. 2(b) shows the eventual SSI of these two frequency-sheared pulses.

TABLE I. Main parameters of the SXFEL.

| Parameters | Value | Unit |
| --- | --- | --- |
| Electron beam energy | 1.5 | GeV |
| Peak current | 800 | A |
| Normalized emittance | 0.65 | mm·mrad |
| Slice energy spread | 150 | keV |
| Main undulator period | 1.6 | cm |

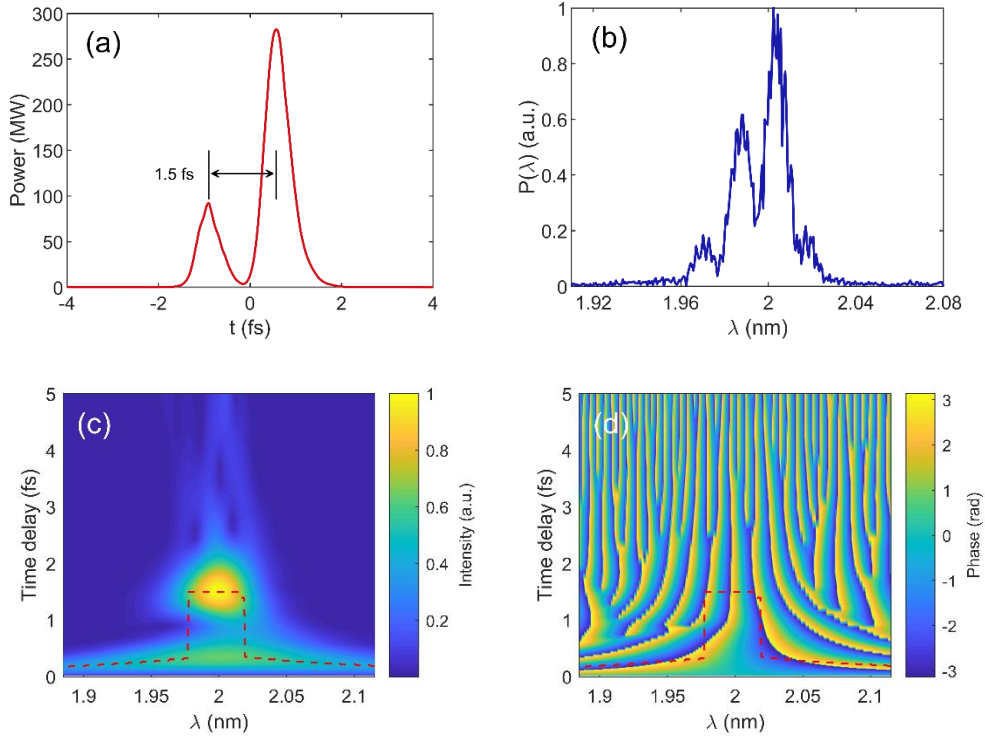

FIG. 2. Simulation and data analysis processes for the proposed method. (a) The temporal profiles of the sample pulse (right) and reference pulse (left). (b) The spectral shearing interferogram of these frequency-sheared pulses. Time-frequency magnitude topography (c) and phase topography (d) after WT.

The next step is to extract the phase information from the SSI. Assuming the FEL pulse with an electric field $\tilde{E}(\omega) = |\tilde{E}(\omega)|\exp[i\phi(\omega)]$, the SSI function will take the following

form:

$$S(\omega,\tau) = |\tilde{E}(\omega)|^2 + |\tilde{E}(\omega+\Omega)|^2 \\ +2|\tilde{E}(\omega)||\tilde{E}(\omega+\Omega)|\cos[\phi(\omega)-\phi(\omega+\Omega)+\omega\tau] \tag{2}$$

where $\Omega$ is the frequency shift and $\tau$ is the time delay between two pulses. As a signal analysis tool, WT technique has great advantages in analyzing signals with complicated frequency components. The WT of SSI function $S(\omega')$ can be expressed as

$$W(\Delta\omega,\omega) = \frac{1}{\Delta\omega}\int_{-\infty}^{+\infty} S(\omega')\psi^*\left(\frac{\omega'-\omega}{\Delta\omega}\right)d\omega', \tag{3}$$

where $\psi[(\omega'-\omega)/\Delta\omega]$ is the daughter wavelet generated by dilation and translation of Gabor mother wavelet $\psi(\omega) = e^{(-\omega^2/2\sigma^2+i2\pi\omega)}/(\sigma^2\pi)^{1/4}$ (here $\sigma = 1/\sqrt{2\ln 2}$). Based on this method, the shaping factor of Gabor wavelet has little effect on the phase retrieval [37]. According to the relationship between $\tau$ and fringe spacing $\Delta\omega$ of the shearing interferogram, i.e., $\tau = 2\pi/\Delta\omega$, Eq. 3 can be written as another form:

$$W(t,\omega) = \frac{t}{2\pi}\int_{-\infty}^{+\infty} S(\omega')\psi^*[\frac{(\omega'-\omega)\cdot t}{2\pi}]d\omega'. \tag{4}$$

After performing the WT of $S(\omega,\tau)$, one-dimensional frequency domain SSI signal is converted into two time-frequency graphics in Fig. 2(c, d): magnitude topography and phase topography, reflecting the intensity and phase information in the time-frequency plane, respectively. We search for the maximum value of magnitude topography at each frequency point, i.e., the red ridge curve in Fig. 2(c), and then project it onto the phase topography, from which the phase $\phi(\omega)-\phi(\omega+\Omega)+\omega\tau$ can be extracted directly. The linear term $\omega\tau$ can be removed utilizing the time delay $\tau$ read from Fig. 2(c). And then the spectral phase $\phi(\omega)$ can be retrieved through an inversion routine [33]. The WT algorithm avoids the influence of phase noise as much as possible since the phase along the maximum value of magnitude topography is extracted from the phase topography. The spectrum of sample pulse can also be recovered from the interferogram. Assuming $S_1(\omega) = |\tilde{E}(\omega)|^2 + |\tilde{E}(\omega+\Omega)|^2$, $S_2(\omega) = 2|\tilde{E}(\omega)||\tilde{E}(\omega+\Omega)|$, the spectral intensity $|\tilde{E}(\omega)|$ can be

expressed as

$$|\tilde{E}(\omega)| = \frac{1}{2}\left[\sqrt{S_1(\omega) + S_2(\omega)} + \sqrt{S_1(\omega) - S_2(\omega)}\right]. \quad (5)$$

With the above method, the spectrum and the retrieved spectral phase were obtained and given in Fig. 3(a). After the Fourier transform, the reconstructed temporal profile and phase are present in Fig. 3(b) (red lines). The duration of reconstructed pulse is about 590 attoseconds, fitting well with the initial pulse duration of 610 attoseconds exported from *GENESIS* (blue lines). In order to verify the stability of the proposed method, multishot simulations have also been performed by varying the initial stochastic noise of the electron beams. Simulation results are summarized in Fig. 3(c). The reconstruction error of the pulse duration is calculated to be about 5.6% (rms), which is acceptable for pulse characterization. The error can be reduced by increasing $\tau$ to create more interference fringes. However, $\tau$ has an upper limit determined by the $R56$ so that the bunching is not blurred after the chicane, i.e., $R56 < \gamma\lambda_s/2\pi\sigma_\gamma$ [42]. Therefore, for shorter pulse durations, the precision of the proposed method will improve due to the larger bandwidths (more fringes).

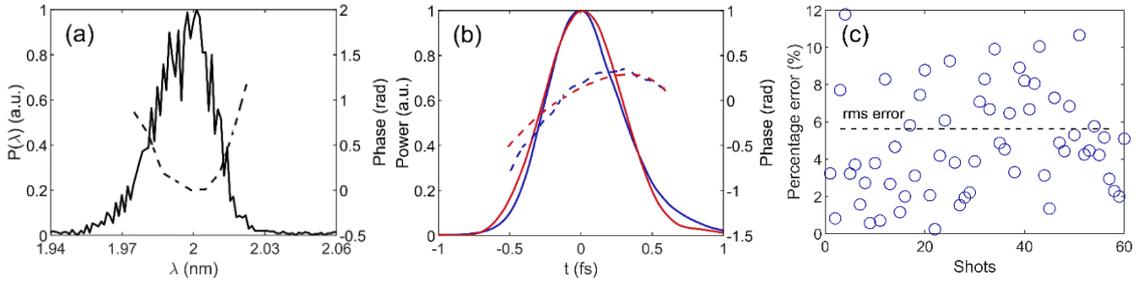

FIG. 3. Spectro-temporal reconstructions of attosecond pulses. (a) Spectrum (solid line) and retrieved spectral phase (dashed line). (b) Temporal profiles and phases of the reconstructed pulse (red) and that directly exported from simulation (blue). (c) Statistical analysis of the reconstructed pulse duration for multishot simulations (60 shots).

*Experiment.* The feasibility of the proposed method has been demonstrated using the experimental data measured at Shanghai Deep Ultraviolet Free Electron Laser test facility (SDUV-FEL) [43,44]. During the experiments, the facility was operated with the high-gain harmonic generation (HGHG) mode [44], as shown in Fig. 4(a). Comparing with SASE, HGHG can produce FEL pulses with stable and repeatable properties, providing ideal conditions for testing the feasibility and reproducibility of the proposed method. A 148 MeV electron beam with bunch charge of 200 pC, bunch length of about 8.8 ps interacts with a 1200 nm seed laser pulse with duration of about 150 fs (FWHM) in the modulator (M1) with period length of 50 mm and total length of 0.5 m to generate coherent energy modulation. After passing through a dispersive chicane (DS1) to convert the energy modulation into microbunching, the electron beam is sent into the radiator (R1) with period length of 40 mm and total length of 1.6 m to generate FEL pulse (sample pulse) at 600 nm ($2^{nd}$ harmonic) with the temporal properties similar to the seed laser. The electron beam then consecutively passages through a small chicane (shifter) to induce a time delay of 600 fs and an afterburner (M2) with period length of 40 mm and total length of 0.64 m to produce the reference pulse. The resonance condition of M2 was tuned to shift the central wavelength of the reference pulse by $\Delta\lambda = 0.55\ nm$, which corresponds to a frequency shift of $\Omega \approx 2.88 \times 10^{12}$ rad. The SSI signal was measured by a spectrometer (TRIAX-550, Jobin Yvon) downstream of the afterburner.

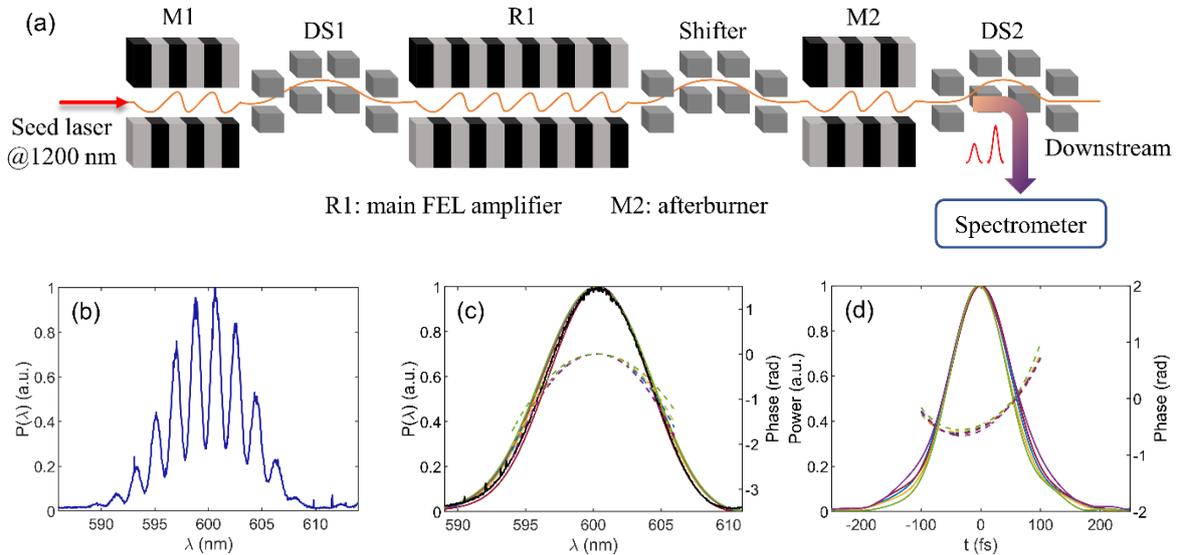

FIG. 4. Experimental layout and results. (a) Layout of the proof-of-principle experiment for the proposed method. (b) Typical spectral shearing interferogram. (c) Directly measured spectrum (black solid line), reconstructed spectra (color solid lines) and spectral phases

(color dashed lines) obtained from 5 consecutive FEL shots. (d) Reconstructed temporal profiles (solid lines) and phases (dashed lines) obtained from 5 consecutive FEL shots. Lines of the same color represent the same FEL pulse.

Fig. 4(b) shows a typical single-shot interferogram detected by the spectrometer. With the proposed method, Fig. 4(c, d) show the reconstruction results in the spectral and temporal domains for five consecutive FEL shots. The retrieved phases and longitudinal profiles are consistent with each other, demonstrating the reproducibility of the proposed method. The spectral envelopes (color solid lines) calculated by Eq. 5 match well with the directly measured single-shot spectrum (black solid line). The retrived temporal phase profiles have similar curvatures that reflect the positive frequency chirp in the seed laser. The average spectral bandwidth and pulse duration of the reconstructed pulses are 8.9 nm and 175 fs (FWHM), respectively, in accordance with the theoretical predictions considering the slippage effect in the radiator. The fluctuation of the retrieved pulse duration is less than 6 %.

*Conclusion.* In conclusion, we have proposed and demonstrated a novel and feasible diagnostic method that can fully characterize the spectro-temporal information of the ultrashort FEL pulse on a single-shot basis. By performing Wavelet-transform of the spectral shearing interferogram, one can retrieve the spectral phase and reconstruct the FEL pulse with attosecond resolution. The proposed method can be easily implemented in current XFEL facilities by adding a small chicane between the last two undulator segments of the FEL amplifier, making the last undulator work as the afterburner and using an online spectrometer to detect the SSI signal. Alternatively, one could use the quadrupole before the afterburner to kicker the electron beam with a tiny angle to split the sample pulse and the reference pulse after a long distance [45], which would allow blocking the reference pulse after detection and minimizing its effects on user experiments. This method would enable us to tune the machine for generating custom-tailored FEL pulses and explore new possibilities of attosecond science with x-ray attosecond pump-probe techniques.

*Acknowledgments.* This work was supported by the National Natural Science Foundation of China (12122514, 11975300) and the Shanghai Science and Technology Committee Rising-Star Program (20QA1410100).

**References**

[1] M. Schultze et al., *Delay in Photoemission*, Science (1979) **328**, 1658 (2010).

[2] P. Eckle, A. N. Pfeiffer, C. Cirelli, A. Staudte, R. Dörner, H. G. Muller, M. Büttiker, and U. Keller, *Attosecond Ionization and Tunneling Delay Time Measurements in Helium*, Science (1979) **322**, 1525 (2008).

[3] E. Goulielmakis et al., *Real-Time Observation of Valence Electron Motion*, Nature **466**, 739 (2010).

[4] P. M. Kraus et al., *Measurement and Laser Control of Attosecond Charge Migration in Ionized Iodoacetylene*, Science (1979) **350**, 790 (2015).

[5] S. Baker, J. S. Robinson, C. A. Haworth, H. Teng, R. A. Smith, C. C. Chirilă, M. Lein, J. W. G. Tisch, and J. P. Marangos, *Probing Proton Dynamics in Molecules on an Attosecond Time Scale*, Science (1979) **312**, 424 (2006).

[6] M. Hentschel, R. Kienberger, Ch. Spielmann, G. A. Reider, N. Milosevic, T. Brabec, P. Corkum, U. Heinzmann, M. Drescher, and F. Krausz, *Attosecond Metrology*, Nature **414**, 509 (2001).

[7] H. Vincenti and F. Quéré, *Attosecond Lighthouses: How To Use Spatiotemporally Coupled Light Fields To Generate Isolated Attosecond Pulses*, Phys Rev Lett **108**, 113904 (2012).

[8] J. Li, J. Lu, A. Chew, S. Han, J. Li, Y. Wu, H. Wang, S. Ghimire, and Z. Chang, *Attosecond Science Based on High Harmonic Generation from Gases and Solids*, Nat Commun **11**, 2748 (2020).

[9] Ch. Spielmann, N. H. Burnett, S. Sartania, R. Koppitsch, M. Schnürer, C. Kan, M. Lenzner, P. Wobrauschek, and F. Krausz, *Generation of Coherent X-Rays in the Water Window Using 5-Femtosecond Laser Pulses*, Science (1979) **278**, 661 (1997).

[10] E. J. Takahashi, T. Kanai, K. L. Ishikawa, Y. Nabekawa, and K. Midorikawa,

*Coherent Water Window X Ray by Phase-Matched High-Order Harmonic Generation in Neutral Media*, Phys Rev Lett **101**, 253901 (2008).

[11] P. Emma et al., *First Lasing and Operation of an Ångstrom-Wavelength Free-Electron Laser*, Nat Photonics **4**, 641 (2010).

[12] E. Allaria et al., *Two-Stage Seeded Soft-X-Ray Free-Electron Laser*, Nat Photonics **7**, 913 (2013).

[13] A. A. Zholents, *Method of an Enhanced Self-Amplified Spontaneous Emission for x-Ray Free Electron Lasers*, Physical Review Special Topics - Accelerators and Beams **8**, 040701 (2005).

[14] J. Duris et al., *Tunable Isolated Attosecond X-Ray Pulses with Gigawatt Peak Power from a Free-Electron Laser*, Nat Photonics **14**, 30 (2020).

[15] Z. Wang, C. Feng, and Z. Zhao, *Generating Isolated Terawatt-Attosecond x-Ray Pulses via a Chirped-Laser-Enhanced High-Gain Free-Electron Laser*, Physical Review Accelerators and Beams **20**, 040701 (2017).

[16] Z. Qi, C. Feng, H. Deng, B. Liu, and Z. Zhao, *Generating Attosecond X-Ray Pulses through an Angular Dispersion Enhanced Self-Amplified Spontaneous Emission Free Electron Laser*, Physical Review Accelerators and Beams **21**, 120703 (2018).

[17] Z. Zhang, J. Duris, J. P. MacArthur, Z. Huang, and A. Marinelli, *Double Chirp-Taper x-Ray Free-Electron Laser for Attosecond Pump-Probe Experiments*, Physical Review Accelerators and Beams **22**, 050701 (2019).

[18] L. Tu, Z. Qi, Z. Wang, S. Zhao, Y. Lu, W. Fan, H. Sun, X. Wang, C. Feng, and Z. Zhao, *Improving the Performance of an Ultrashort Soft X-Ray Free-Electron Laser via Attosecond Afterburners*, Applied Sciences **12**, 11850 (2022).

[19] D. Xiang, Z. Huang, and G. Stupakov, *Generation of Intense Attosecond X-Ray Pulses Using Ultraviolet Laser Induced Microbunching in Electron Beams*, Physical Review Special Topics - Accelerators and Beams **12**, 060701 (2009).

[20] T. Tanaka, *Proposal to Generate an Isolated Monocycle X-Ray Pulse by Counteracting the Slippage Effect in Free-Electron Lasers*, Phys Rev Lett **114**,

044801 (2015).

[21] C. Feng, J. Chen, and Z. Zhao, *Generating Stable Attosecond X-Ray Pulse Trains with a Mode-Locked Seeded Free-Electron Laser*, Physical Review Special Topics - Accelerators and Beams **15**, 080703 (2012).

[22] Y. Ding, C. Behrens, P. Emma, J. Frisch, Z. Huang, H. Loos, P. Krejcik, and M.-H. Wang, *Femtosecond X-Ray Pulse Temporal Characterization in Free-Electron Lasers Using a Transverse Deflector*, Physical Review Special Topics - Accelerators and Beams **14**, 120701 (2011).

[23] C. Behrens et al., *Few-Femtosecond Time-Resolved Measurements of X-Ray Free-Electron Lasers*, Nat Commun **5**, (2014).

[24] C. Feng et al., *Coherent and Ultra-Short Soft X-Ray Pulses from Echo-Enabled Harmonic Cascade Free-Electron Laser*, Optica (2022).

[25] T. Plath, C. Lechner, V. Miltchev, P. Amstutz, N. Ekanayake, L. L. Lazzarino, T. Maltezopoulos, J. Bödewadt, T. Laarmann, and J. Roßbach, *Mapping Few-Femtosecond Slices of Ultra-Relativistic Electron Bunches*, Sci Rep **7**, 2431 (2017).

[26] E. Allaria et al., *Energy Slicing Analysis for Time-Resolved Measurement of Electron-Beam Properties*, Physical Review Special Topics - Accelerators and Beams **17**, 010704 (2014).

[27] P. Radcliffe et al., *Single-Shot Characterization of Independent Femtosecond Extreme Ultraviolet Free Electron and Infrared Laser Pulses*, Appl Phys Lett **90**, (2007).

[28] V. S. Yakovlev, J. Gagnon, N. Karpowicz, and F. Krausz, *Attosecond Streaking Enables the Measurement of Quantum Phase*, Phys Rev Lett **105**, 073001 (2010).

[29] W. Helml et al., *Measuring the Temporal Structure of Few-Femtosecond Free-Electron Laser X-Ray Pulses Directly in the Time Domain*, Nat Photonics **8**, 950 (2014).

[30] I. Grguraš et al., *Ultrafast X-Ray Pulse Characterization at Free-Electron Lasers*, Nat Photonics **6**, 852 (2012).

[31] S. Li et al., Attosecond Coherent Electron Motion in Auger-Meitner Decay, 2022.

[32] N. Hartmann et al., *Attosecond Time–Energy Structure of X-Ray Free-Electron Laser Pulses*, Nat Photonics **12**, 215 (2018).

[33] C. Iaconis and I. A. Walmsley, *Self-Referencing Spectral Interferometry for Measuring Ultrashort Optical Pulses*, IEEE J Quantum Electron **35**, 501 (1999).

[34] M. E. Anderson, L. E. E. de Araujo, E. M. Kosik, and I. A. Walmsley, *The Effects of Noise on Ultrashort-Optical-Pulse Measurement Using SPIDER*, Applied Physics B **70**, S85 (2000).

[35] S. Jensen and M. E. Anderson, *Measuring Ultrashort Optical Pulses in the Presence of Noise: An Empirical Study of the Performance of Spectral Phase Interferometry for Direct Electric Field Reconstruction*, Appl Opt **43**, 883 (2004).

[36] G. de Ninno et al., *Single-Shot Spectro-Temporal Characterization of XUV Pulses from a Seeded Free-Electron Laser*, Nat Commun **6**, 8075 (2015).

[37] Y. Deng, Z. Wu, S. Cao, L. Chai, C. yue Wang, and Z. Zhang, *Spectral Phase Extraction from Spectral Shearing Interferogram for Structured Spectrum of Femtosecond Optical Pulses*, Opt Commun **268**, 1 (2006).

[38] Y. Deng, Z. Wu, L. Chai, C.-Y. Wang, K. Yamane, R. Morita, M. Yamashita, and Z. Zhang, Wavelet-Transform Analysis of Spectral Shearing Interferometry for Phase Reconstruction of Femtosecond Optical Pulses, 2005.

[39] E. Allaria, G. de Ninno, and C. Spezzani, *Experimental Demonstration of Frequency Pulling in Single-Pass Free-Electron Lasers*, Opt Express **19**, 10619 (2011).

[40] S. Reiche, *GENESIS 1.3: A Fully 3D Time-Dependent FEL Simulation Code*, Nucl Instrum Methods Phys Res A **429**, 243 (1999).

[41] B. Liu et al., *The Sxfel Upgrade: From Test Facility to User Facility*, Applied Sciences (Switzerland) **12**, (2022).

[42] E. A. Schneidmiller, *Application of a Modified Chirp-Taper Scheme for Generation of Attosecond Pulses in Extreme Ultraviolet and Soft x-Ray Free Electron Lasers*, Physical Review Accelerators and Beams **25**, 010701 (2022).

[43] Z. T. Zhao et al., *First Lasing of an Echo-Enabled Harmonic Generation Free-Electron Laser*, Nat Photonics **6**, 360 (2012).

[44] C. Feng, M. Zhang, G. Q. Lin, Q. Gu, H. X. Deng, J. H. Chen, D. Wang, and Z. T. Zhao, *Design Study for the Cascaded HGHG Experiment Based on the SDUV-FEL*, Chinese Science Bulletin **57**, 3423 (2012).

[45] J. P. MacArthur, A. A. Lutman, J. Krzywinski, and Z. Huang, *Microbunch Rotation and Coherent Undulator Radiation from a Kicked Electron Beam*, Phys Rev X **8**, 041036 (2018).